\documentclass{aa}
\usepackage[round]{natbib}
\usepackage{graphicx, amsmath, aas_macros}
\usepackage{amssymb}
\usepackage[scriptsize]{subfigure}
\usepackage{url}
\usepackage[scriptsize]{subfigure}

\begin{document}

   \title{VLBI observations of weak sources using fast frequency switching}

   \author{E. Middelberg\thanks{Present address: Australia Telescope National Facility, 
           PO Box 76, Epping NSW 1710, Australia},
          \inst{1}
          A. L. Roy,
          \inst{1,2}
          R. C. Walker,
          \inst{3}
          \and
          H. Falcke
          \inst{4}
         }

   \institute{Max-Planck-Institut f\"ur Radioastronomie, Auf dem H\"ugel 69, D-53121 Bonn, Germany\\
              \email{enno.middelberg@csiro.au, aroy@mpifr-bonn.mpg.de}
         \and
              Geod\"atisches Institut der Universit\"at Bonn, Nussallee 17, D-53115 Bonn, Germany
         \and
              National Radio Astronomy Observatory, 
              P.O. Box 0, Socorro, NM, 87801, USA\\
              \email{cwalker@aoc.nrao.edu}
         \and
              ASTRON, P.O. Box 2, 7990 AA Dwingeloo, The Netherlands\\
              \email{falcke@astron.nl}
}
   \offprints{E. Middelberg}

\abstract{We have developed a new phase referencing technique for high
frequency VLBI observations. In conventional phase referencing, one
interleaves short scans on a nearby phase calibrator between the
target source observations. In fast frequency switching described
here, one observes the target source continuously while switching
rapidly between the target frequency and a lower reference
frequency. We demonstrate that the technique allows phase calibration
almost reaching the thermal noise limit and present the first
detection of the AGN in the FR\,I radio galaxy
NGC\,4261 at 86 GHz. Although point-like, this is the
weakest source ever detected with VLBI at this frequency.
\\
 }

\authorrunning{Middelberg et al.}
\titlerunning{Fast frequency switching}

   \maketitle

\section{Introduction}

The regions where jets from active galactic nuclei (AGN) are launched
and collimated are difficult to observe with VLBI because they lie
very close to the black hole and most bright objects are very
distant. Thus, there are still very few observational constraints on
jet formation. Only in the closest AGN can the highest resolution
observations resolve several tens of Schwarzschild radii ($R_{\rm
s}$), comparable to the scale of $10\,R_{\rm s}$ to $1000\,R_{\rm s}$
where jet formation is expected to take place (e.g.,
\citealt{Koide2000},
\citealt{Appl1993}).

One of the best targets for resolving details in the jet is NGC\,4261
(3C\,270). It is an elliptical low-luminosity FR\,I radio galaxy
hosting a $4.9\times10^8\,M_{\odot}$ black hole
(\citealt{Ferrarese1996}) which powers a double-sided radio jet
(\citealt{Jones1997, Jones2000, Jones2001}).  Given the distance to
NGC\,4261 of 28.2\,Mpc ($H_0={\rm 75\,km\,s^{-1}\,Mpc^{-1}}$) and its
black hole mass, 86\,GHz VLBI observations are expected to resolve
$200\,R_{\rm s}$. This measurement would be, after diameter
measurements of \object{Sgr\,A*} ($24\,R_{\rm s}$,
\citealt{Bower2004}, $17\,R_{\rm s}$,
\citealt{Krichbaum1998}) and \object{M\,87} ($32\,R_{\rm s}$,
\citealt{Junor1999}, $46\,R_{\rm s}$, \citealt{Ly2004})  the
third-highest resolution image ever achieved in terms of Schwarzschild
radii.  Furthermore, the high inclination of the jets makes
NGC\,4261 a good candidate to look for a core-shift with
frequency. Core-shift measurements are of considerable astrophysical
interest. They can help to discriminate between the
conically-expanding jet model (\citealt{Blandford1979}) and
advection-dominated accretion flows (ADAFs, e.g.,
\citealt{Narayan1998}) in low-luminosity AGN, to find the true
location of the AGN, to determine the jet magnetic field strength and
to test for potential foreground absorbers (\citealt{Lobanov1998}).
Although NGC\,4261 therefore is an interesting source to
study the jet formation process, it is unfortunately not possible to
directly observe NGC\,4261 with 86\,GHz VLBI, because it is
too weak. This paper demonstrates a new calibration method that makes
such an observation possible.\\

High frequency observations involve a number of serious problems: the
sources are usually weak because the emissivity of optically thin
synchrotron sources drops as $\nu^{-0.7}$, the aperture efficiencies
of most of the radio telescopes drop to 15\,\% or less because their
surface accuracies were specified for cm-wavelength operation, the
receiver performances become disproportionately worse because of
higher amplifier noise, the atmospheric contribution to the system
temperatures increases towards higher frequency, and the atmospheric
coherence that limits the integration time decreases as $1/\nu$.

However, the integration times could be prolonged if the atmospheric
phase fluctuations could be calibrated. This can be done using
self-calibration, or if the source is too weak, then using phase
referencing (e.g., \citealt{Shapiro1979, Marcaide1984, Alef1988}).
Both are established calibration techniques and phase referencing can
be used to lower detection thresholds to the sub-mJy level and can
accurately determine source positions. In phase referencing, a strong
calibrator source is observed frequently (every few minutes, depending
on observing frequency) to calibrate the visibility phases of the
target source integrations, i.e., the telescopes cycle between the
target source and the phase calibrator source. In millimetre VLBI the
technique is not commonly used because of the need for a suitable,
strong phase calibrator in the vicinity of the target source and from
a combination of short atmospheric coherence time and relatively long
telescope slewing times. However, a successful proof of concept exists
for 86\,GHz VLBI observations (\citealt{Porcas2002}).

In this paper, we describe a novel way of calibrating visibility
phases to overcome these limitations, using interleaved observations
at a low and a high frequency.

\section{Principle of phase correction}
\label{sec:principle}

\subsection{Self-calibration}

The phase of the complex visibility function which an interferometer
measures is altered by instrumental drifts, antenna and source
position errors, and path length changes in the earth's
atmosphere. These errors can be accounted for using self-calibration,
which is similar to adaptive optics in optical astronomy. Starting
with a point source model in the field centre, and refining the model
iteratively, antenna-based correction phases are derived that make the
visibility phases compliant with the model. Similar to adaptive
optics, where a sufficiently bright guide star is required to detect
the shape of the wavefront, self-calibration requires a minimum
signal-to-noise ratio (SNR) on each baseline. Visibilities with an SNR
of less than about five within the atmospheric coherence time
are commonly regarded as non-detections.

\subsection{Scaling and interpolation}

\begin{figure}[ht!]
\centering
 \includegraphics[width=0.8\linewidth, angle=270]{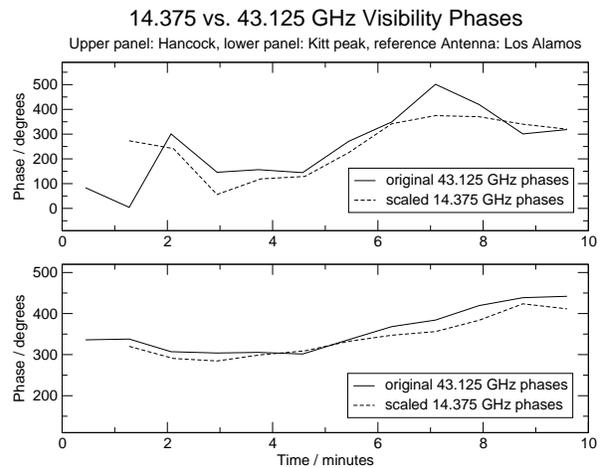}
 \caption{Demonstration of the scalability of phase solutions. 43\,GHz
 fringe-fitted phase solutions (solid lines) on \object{3C\,273}
 compared to 15\,GHz fringe-fitted phase solutions multiplied by the
 frequency ratio (dashed lines) from the VLBA antennas at Hancock
 (upper panel) and Kitt Peak (lower panel) to the reference VLBA
 antenna at Los Alamos. The phases follow each other very well.}
 \label{fig:phase}
\end{figure}

One can self-calibrate the visibility phases at one frequency and use
the solutions to calibrate visibilities at another frequency after
multiplying the phases by the frequency ratio, $r$.  This works since
the main source of phase noise in VLBI observations at frequencies
higher than about 5\,GHz is turbulence in the troposphere causing
refractive inhomogeneities which are non-dispersive.  One requires
that the lag between the two measurements does not exceed half the
atmospheric coherence time (Fig.~\ref{fig:phase}).  Instead of using a
phase-referencing calibrator source, the source is phase-referenced to
itself at a lower frequency.

This is possible with the Very Long Baseline Array\footnote{The VLBA is
an instrument of the National Radio Astronomy Observatory, a facility
of the National Science Foundation, operated under cooperative
agreement by Associated Universities, Inc.} (VLBA,
\citealt{Napier1994}) because frequency changes need only a few
seconds, in which the subreflector is moved from one feed horn to the
other, and because the local oscillator phase returns to its original
setting after frequency switching. After multiplying the phase
solutions by the frequency ratio and applying them to the
target-frequency phases, there remains a constant phase offset,
$\Delta\Phi$, between the signal paths at the two frequencies, which
must be calibrated. This can be monitored with frequent observations
of achromatic, strong calibrators, and must be subtracted from the
high-frequency visibility phase.\\

The observed visibility phases using this calibration scheme can be
described as

\begin{equation}
\begin{split}
\phi_{\rm r}^{\rm obs}(t_1) =
  \phi_{\rm r}^{\rm tru}(t_1)+
  \phi_{\rm r}^{\rm ins}(t_1)  
&+\phi_{\rm r}^{\rm pos}(t_1)+
  \phi_{\rm r}^{\rm ant}(t_1)\\
&+\phi_{\rm r}^{\rm tro}(t_1)+
  \phi_{\rm r}^{\rm ion}(t_1)\\
\phi_{\rm t}^{\rm obs}(t_2) =
  \phi_{\rm t}^{\rm tru}(t_2)+
  \phi_{\rm t}^{\rm ins}(t_2) 
&+\phi_{\rm t}^{\rm pos}(t_2)+
  \phi_{\rm t}^{\rm ant}(t_2)\\
&+\phi_{\rm t}^{\rm tro}(t_2)+
  \phi_{\rm t}^{\rm ion}(t_2)\\
\phi_{\rm r}^{\rm obs}(t_3) =
  \phi_{\rm r}^{\rm tru}(t_3)+
  \phi_{\rm r}^{\rm ins}(t_3)  
&+\phi_{\rm r}^{\rm pos}(t_3)+
  \phi_{\rm r}^{\rm ant}(t_3)\\
&+\phi_{\rm r}^{\rm tro}(t_3)+
  \phi_{\rm r}^{\rm ion}(t_3).\\
\end{split}
\label{eq:ffs-phase1}
\end{equation}

Here, $\phi^{\rm tru}$ indicates true visibility phases of the source,
$\phi^{\rm ins}$ is the residual instrumental phase error, $\phi^{\rm
pos}$ and $\phi^{\rm ant}$ are geometric errors arising from source
and antenna position errors, and $\phi^{\rm tro}$ and $\phi^{\rm ion}$
are tropospheric and ionospheric phase noise contributions. The
subscript indices r and t indicate quantities that belong to the
reference frequency, $\nu_{\rm r}$, and the target frequency,
$\nu_{\rm t}$, respectively.

Self-calibration at the reference frequency is used to obtain a source
model and hence to obtain $\phi_{\rm r}^{\rm tru}(t_1)$ and $\phi_{\rm
r}^{\rm tru}(t_3)$.  Self-calibration of the visibilities at the
reference frequency after subtracting this model yields $\phi_{\rm
r}^{\rm cor}$, the sum of $\phi_{\rm r}^{\rm ins}$, $\phi_{\rm r}^{\rm
pos}$, $\phi_{\rm r}^{\rm ant}$, $\phi_{\rm r}^{\rm tro}$ and
$\phi_{\rm r}^{\rm ion}$. This sum is linearly interpolated to the
times where the target frequency was observed, and is scaled by the
frequency ratio, $r$:

\begin{equation}
 r\tilde{\phi}_{\rm r}^{\rm cor}= 
r(\tilde{\phi}_{\rm r}^{\rm ins}+
  \tilde{\phi}_{\rm r}^{\rm pos}+
  \tilde{\phi}_{\rm r}^{\rm ant}+
  \tilde{\phi}_{\rm r}^{\rm tro}+
  \tilde{\phi}_{\rm r}^{\rm ion}),
\end{equation}

where a tilde denotes quantities linearly interpolated to time
$t_2$. The difference between the target frequency visibility phase
and the interpolated and scaled reference frequency correction phase
at time $t_2$ is then

\begin{equation}
\begin{split}
\phi_{\rm t}^{\rm obs}-r\tilde{\phi}_{\rm r}^{\rm cor}= 
\phi_{\rm t}^{\rm tru}
&+(\phi_{\rm t}^{\rm ins}-r\tilde{\phi}_{\rm r}^{\rm ins})
 +(\phi_{\rm t}^{\rm pos}-r\tilde{\phi}_{\rm r}^{\rm pos})\\
&+(\phi_{\rm t}^{\rm ant}-r\tilde{\phi}_{\rm r}^{\rm ant})  
 +(\phi_{\rm t}^{\rm tro}-r\tilde{\phi}_{\rm r}^{\rm tro})\\
&+(\phi_{\rm t}^{\rm ion}-r\tilde{\phi}_{\rm r}^{\rm ion}).
\end{split}
\label{eq:ffs-phase3}
\end{equation}

The instrumental phase offset, $\Delta\Phi=(\phi_{\rm t}^{\rm
ins}-r\tilde{\phi}_{\rm r}^{\rm ins})$, is constant with time and can
be determined from calibrator observations, and so it is known and can
be removed. The antenna position error, like all geometric errors,
scales with frequency and hence $(\phi_{\rm t}^{\rm
ant}-r\tilde{\phi}_{\rm r}^{\rm ant})=0$. A frequency-dependent core
shift, however, as predicted by conical jet models (e.g.,
\citealt{Lobanov1998}), modulates $(\phi_{\rm t}^{\rm
pos}-r\tilde{\phi}_{\rm r}^{\rm pos})$ on each baseline with a
sinusoid. The period of the sinusoid is $23^{\rm h}56^{\rm m}$, its
amplitude depends on the magnitude of the shift and its phase depends
on the direction of the shift relative to the baseline direction. It
is zero only in the absence of a core shift, and therefore provides
useful structural information. The tropospheric phase errors also
scale with frequency, and hence $(\phi_{\rm t}^{\rm
tro}-r\tilde{\phi}_{\rm r}^{\rm tro})=0$.  The ionospheric phase
errors do not scale linearly with frequency and so cannot be removed
using this technique, requiring that they be measured separately. We
assume for the moment that this has been done, but see the next
section for the impact of ionospheric effects on the data presented
here. The remaining terms describe the difference between the target
frequency phase and the scaled and interpolated reference frequency
phase corrections as the target frequency visibility phase plus the
position offset:

\begin{equation}
\phi_{\rm t}^{\rm obs}-r\tilde{\phi}_{\rm r}^{\rm cor}= 
\phi_{\rm t}^{\rm tru}
+(\phi_{\rm t}^{\rm pos}-r\tilde{\phi}_{\rm r}^{\rm pos}).
\label{eq:ffs-phase4}
\end{equation}

Thus, the true high-frequency visibilities are phase-referenced to the
source's low-frequency visibilities, and so the technique can prolong
coherence and can measure the position shift of cores in AGN with
frequency.

\subsection{Ionospheric path length changes}

For illustration, we assumed in the previous section that the
ionospheric effects were calibrated. Ionospheric path length changes
can probably be determined using interspersed, wide-band scans at a
low frequency, e.g., in the 1.4\,GHz band, where the effect is
strong. We had not appreciated before observing that the ionosphere
would still be significant at 15\,GHz and had not planned 1.4\,GHz
scans to measure ionosphere. Hence, the high-frequency visibilities
from the observations presented here contained unmodelled ionospheric
path length changes which limited the coherence to half an hour in the
worst case. To remove the remaining long-term phase drifts and
remaining phase offsets required one extra step of self-calibration.
This step of self-calibration loses source position information, which
prevented a core-shift measurement from being made. Still the
extension from the 30\,s atmospheric coherence time to a coherent
integration time of 30\,min yielded a large sensitivity improvement and
allows weaker sources to be detected.\\

A similar observing strategy has been developed by \cite{Kassim1993}
for the VLA, who used scaled phase solutions from 330\,MHz to
calibrate simultaneously observed 74\,MHz data. In this case, the
dominant source of phase errors was ionospheric path length
changes. Using their new calibration technique, they were able to
increase the coherent integration time from $<$\,1\,min to
$>$\,10\,min and to make 74\,MHz images of several radio sources. Fast
frequency switching is also being considered as a standard calibration
mode for the Atacama Large Millimeter Array (ALMA) in the future
(\citealt{DAddario2003}).

\section{Observations}
\label{sec:observations}

\begin{table}[htpb!]
\center
\begin{tabular}{lcc|lcc}
\hline
\hline
Source   & Duration  & Freq.  & 3C\,279    &  5   & 43 \\ 
         & (min)     & pair   & 3C\,273    &  5   & 43 \\ 
(1)      & (2)       & (3)    & 3C\,273    &  5   & 1  \\ 
\cline{1-3}
OJ\,287    & 10   & 15-86  & NGC\,4261  & 25   & 15-43  \\  
OJ\,287    & 10   & 15-43  & 3C\,273    &  5   & 15-43  \\  
NGC\,4261  & 25   & 15-43  & 3C\,273    &  5   & 15-86  \\  
3C\,273    &  5   & 15-43  & NGC\,4261  & 25   & 15-86  \\  
NGC\,4261  & 25   & 15-86  & 3C\,279    &  5   & 15-43  \\  
3C\,273    &  5   & 15-86  & 3C\,279    &  5   & 15-86  \\  
3C\,273    &  5   & 15-43  &            &      &        \\  
           &      &        & 3C\,273    &  10  & 15-43  \\  
OJ\,287    &  5   & 15$^a$ &            &      & 15-    \\  
OJ\,287    &  5   & 43$^a$ & NGC\,4261  & 25   & 15-43  \\  
OJ\,287    &  5   & 86$^a$ & 3C\,273    &  5   & 15-43  \\  
           &      &        & 3C\,273    &  5   & 15-86  \\  
NGC\,4261  & 25   & 15-43  & NGC\,4261  & 25   & 15-86  \\  
3C\,279    &  5   & 15-43  & 3C\,279    &  5   & 15-86  \\  
3C\,279    &  5   & 15-86  & 3C\,279    &  5   & 15-43  \\  
3C\,273    &  5   & 15-86  &            &      &        \\  
3C\,273    &  5   & 15-43  & 3C\,279    &  5   & 15$^a$ \\  
NGC\,4261  & 25   & 15-43  & 3C\,279    &  5   & 43$^a$ \\  
3C\,273    &  5   & 15-43  & 3C\,279    &  5   & 86$^a$ \\  
3C\,273    &  5   & 15-86  &            &      &        \\  
NGC\,4261  & 25   & 15-86  & 3C\,273    &  5   & 15-43  \\  
3C\,279    &  5   & 15-86  & 3C\,273    &  5   & 15-86  \\  
3C\,279    &  5   & 15-43  & NGC\,4261  & 25   & 15-86  \\  
           &      &        & 3C\,273    &  5   & 15-86  \\  
3C\,273    & 10   & 15-43  & 3C\,273    &  5   & 15-43  \\  
           &      &        & NGC\,4261  & 25   & 15-43  \\  
NGC\,4261  & 25   & 15-43  & 3C\,279    &  5   & 15-43  \\  
3C\,273    &  5   & 15-43  & 3C\,279    &  5   & 15-86  \\  
3C\,273    &  5   & 15-86  & 3C\,273    &  5   & 15-43  \\  
NGC\,4261  & 25   & 15-86  &            &      &        \\  
3C\,279    &  5   & 15-86  & 3C\,345    &  5   & 15$^a$ \\  
           &      &        & 3C\,345    &  5   & 43$^a$ \\  
3C\,279    &  5   & 15$^a$ & 3C\,345    &  5   & 86$^a$ \\  
3C\,279    &  5   & 43$^a$              &      &        \\  
3C\,279    &  5   & 86$^a$ & 3C\,345    &  10  & 15-43  \\  
           &      &        & 3C\,345    &  10  & 15-86  \\  
\hline
\end{tabular}
\caption{Summary of observations May 5, 2003. The sources observed are
in column (1), the scan duration is in column (2) and the frequency
pairs in GHz are in column (3). An $a$ indicates fringe finder scans
during which no frequency switching was done.}
\label{tab:scans}
\end{table}

We observed NGC\,4261 on May 5, 2003 with the ten VLBA
stations and we interleaved observations of \object{3C\,273} and
\object{3C\,279} to calibrate the inter-frequency offset and to
test the technique on strong sources.  Dynamic scheduling allowed us
to observe during a period of good to excellent weather at most
stations, using 256\,Mbps to record a bandwidth of 64\,MHz with 2-bit
sampling. We recorded LCP only, the data were divided into 8 IFs with
8\,MHz bandwidth, each of which were subdivided into 64 spectral
channels 125\,kHz wide. The correlator integration time was one second
to allow monitoring of the phases with high time resolution. All
antennas performed well, except for Fort Davis, where a receiver
problem caused complete loss of 43\,GHz data. A summary of this
observing run is given in Table~\ref{tab:scans}.

In this paper, two integrations at $\nu_{\rm r}$ and $\nu_{\rm t}$
are called a ``cycle'', each integration of which is called a
``half-cycle'', and a sequence of cycles on the same source is called
a ``scan''. Long (several minutes), continuous integrations on a
single source at one frequency, e.g. fringe-finder observations, are
also called ``scans''.

Several considerations influenced the experiment design.

\subsection{Frequency choice} The target frequency should be an integer 
multiple of the reference frequency to avoid having to unwrap phase
wraps. For example, if the frequency ratio, $r$, is the non-integer
value of 2.5, and the reference frequency phase wraps from
$359^{\circ}$ to $0^{\circ}$, then the scaled target frequency phase
will jump from $897.5^{\circ}$ ($=177.5^{\circ}$) to $0^{\circ}$,
introducing a phase jump of $177.5^{\circ}$ into the calibration phase
for a $1^{\circ}$ phase change at $\nu_{\rm r}$. In contrast,
choosing $r=2.0$, when the reference frequency phase wraps from
$359^{\circ}$ to $0^{\circ}$, the scaled target frequency phase
changes from $718^{\circ}$ ($=358^{\circ}$) to $0^{\circ}$,
corresponding to a change of $2^{\circ}$ for a $1^{\circ}$ phase
change at $\nu_{\rm r}$. Hence, for integer values of $r$, phase
wraps at the reference frequency introduce integer multiples of
$360^{\circ}$ at the target frequency and so have no effect.

We chose a reference frequency of 14.375\,GHz since the third and
sixth harmonics at 43.125\,GHz and 86.25\,GHz lie within the VLBA
receiver bands. For convenience, we will refer to these frequencies as
``15\,GHz'', ``43\,GHz'' and ``86\,GHz'', respectively. These
frequencies should then be shifted slightly to end in .49\,MHz or
.99\,MHz to allow correct operation of the VLBA pulse calibration
detection system, and the final two digits should not change during
the experiment as a change causes a time-consuming reconfiguration of
the formatter.  Thus, optimal frequencies are 14.37499\,GHz,
43.12499\,GHz and 86.24999\,GHz.

\subsection{Integration times} We chose a cycle time of 50\,s, of
which 22\,s were spent at the reference frequency of 15\,GHz and the
remaining 28\,s were spent at the target frequency, either 43\,GHz or
86\,GHz.  An average time of 7\,s per half-cycle was lost in moving
the subreflector between the feed horns, resulting in net integration
times of 15\,s at $\nu_{\rm r}$ and 21\,s at $\nu_{\rm t}$. The
integration times are a compromise, depending on source brightness,
antenna sensitivity and expected weather conditions. This setup
yielded a $5\,\sigma$ detection limit of 89\,mJy in 15\,s at 15\,GHz
for the VLBA on a single baseline.

\subsection{Calibrator scans} Calibrators must be observed frequently to
monitor the phase offset $\Delta\Phi=\phi_{\rm t}^{\rm
ins}-r\tilde{\phi}_{\rm r}^{\rm ins}$ between the two frequencies. A
constraint which is important for core-shift measurements is that the
calibrators must be achromatic, i.e., they must not have their own
frequency-dependent core shifts. In the case of the observations
presented here, uncalibrated ionospheric errors prevented such a
measurement of $\Delta\Phi$. We used self-calibration with a long
solution interval to calibrate the residual phase errors, and did not
require measurement of the instrumental phase offset on calibrators.

As the core-shift information is not generally available, the best
strategy is to observe at least two calibrators to test for a possible
core shift in the calibrators. We included adjacent scans on two
different calibrators to measure $\Delta\Phi$ before and after target
source observations. Five of the calibrator scans were twice as long
to provide more time for tests with strong signals.

\section{Data reduction}
\label{sec:calibration}

\subsection{Standard steps}

Data reduction was carried out in AIPS. A calibration table entry was
generated every 4.8\,s to provide high temporal resolution. The
amplitudes were calibrated using $T_{\rm sys}$ and gain measurements
provided by automatic noise-adding radiometry, and amplitude
corrections for errors in the sampler thresholds were performed using
autocorrelation data. Phase corrections for parallactic angles were
applied (this step is not strictly required for the method to work)
and a simple bandpass correction was derived at each frequency from
one of the fringe finder scans.\\

The VLBA's pulse calibration system did not deliver data to calibrate
the phase offsets between the IFs because the frequencies were changed
too quickly. The pulse calibration system has a default integration
time of 10\,s, plus one second for readout. If any part of that
integration time is during a part of the scan that is flagged by the
online system, then the integration is not accepted. Thus, one loses
the first two integrations because the online system conservatively
flags 10\,s to 11\,s after a frequency change, and there is an overlap
between the flagged time and the second integration. During the whole
experiment, only $\sim10$ useful pulse-cal measurements were recorded
per station at 15\,GHz (half-cycle time 22\,s), and $\sim100$ at
43\,GHz (half-cycle time 28\,s), compared to the number of half-cycles
at these frequencies of $\sim630$ and $\sim230$, respectively. We
therefore used fringe-fitting on the same fringe finder scans as were
used for bandpass calibration to correct for instrumental delays and
inter-IF phase offsets. These offsets were found to be stable over the
experiment.\\

\subsection{Ionospheric correction}

Although the frequencies used in the project are quite high,
ionospheric effects can not be neglected and will prevent a successful
phase transfer if uncorrected. A typical ionospheric delay at a
frequency of $\nu_1=100$\,MHz is 0.1\,$\mu$s (10 turns of phase), but
can be up to $10\,\mu{\rm s}$. The exact number strongly depends on
the time of day, time of year and time in the solar cycle. The delay
scales with frequency as $\nu^{-2}$, and so for a delay at 100\,MHz of
0.1\,$\mu$s, the ionospheric delays at 14.37499\,GHz, 43.12499\,GHz
and 86.24999\,GHz are 4.84\,ps, 0.54\,ps and 0.13\,ps, respectively,
corresponding to $25^\circ$, $8.4^\circ$ and $4.0^\circ$ of phase. The
linear phase versus frequency scaling law used by fast frequency
switching cannot correct phase changes that are induced by the
ionosphere. Ionospheric phase changes have much longer time-scales
than tropospheric changes, and they can be calibrated before
fringe-fitting when the electron content of the ionosphere along the
line of sight is known.  The AIPS task TECOR can use maps of
ionospheric total electron content (TEC) derived from GPS data to
calculate phase and delay corrections. Unfortunately, the error in
these maps can be quite high, up to 20\,\% when the TEC is as high as
a few tens of TEC units (1\,TEC unit equals $10^{16}\,{\rm
electrons\,m^{-2}}$), and up to 50\,\% or higher when the TEC is of
the order of a few TEC units. We have used the TEC maps produced by
the Center for Orbit Determination in Europe
(CODE\footnote{\url{http://www.aiub.unibe.ch/ionosphere.html}}) to
calibrate the effects of the ionosphere. We found that these maps
yielded slightly better results than those from the Jet Propulsion
Laboratory (JPL), i.e., the residual phase errors after scaling were
smaller. We do not know whether this finding is coincidental or
whether the CODE maps generally give better results.

\subsection{Effect of frequency changes on the phases}

Amplitudes needed 2\,s to 3\,s longer to reach their final values
during a frequency change than did the phases. Thus, the visibility
phases are not very sensitive to even large errors in subreflector
rotational position. The repeatability of the phases from cycle to
cycle shows that positioning the subreflector along the optical axis
is repeatable to $<5^{\circ}$ of phase at 15\,GHz, which is much less
than other sources of phase error in fast frequency switching. Six to
seven seconds at the beginning of each half-cycle, when the
subreflector was moved between feed horns, needed to be flagged.\\


\subsection{Phase solution and phase scaling}

We derived phase solutions at 15\,GHz using fringe-fitting, which is
equivalent to self-calibration except that it also derives delay
solutions. Its sensitivity therefore is not as good as self-cal, which
performs a phase search only and is preferred. However, fringe-fitting
yielded good results and a high detection rate in our experiment.

We fringe-fitted the 15\,GHz data using the AIPS task FRING, with an
SNR threshold of 5 and delay and rate search windows of 20\,ns and
50\,mHz, respectively.  As NGC\,4261 has an extended,
double-sided jet structure at 15\,GHz, we made a 15\,GHz image which
we used as a source model in a second run of FRING, so that the phase
solutions did not contain structural phase contributions. The solution
interval was set to 1\,min, yielding one phase, delay and rate
solution per half cycle.  The detection rate was $\sim90\,\%$. The
15\,GHz solution (SN) table was written to a text file with TBOUT to
do the phase scaling outside AIPS. We have written a Python program
(FFSTG, the Fast Frequency Switching Table Generator) that processes
an SN table in the following way.\\

First, a series of timestamps is generated from each pair of
consecutive entries in the input table such that they coincide with
the $\nu_{\rm t}$ half-cycles. Second, from each pair of consecutive
$\nu_{\rm r}$ phase solutions, a solution is interpolated for the new
timestamps consisting of a phase, a phase rate derived from the
$\nu_{\rm r}$ phase solutions and the time interval between the
$\nu_{\rm r}$ scans, and an interpolated delay. Both the phase rate
and the delay do not need to be scaled by $r$ because the rate is
stored in a frequency-independent format (in units of ${\rm
s\,s^{-1}}$) and the delay is non-dispersive. Third, the interpolated
phase solution is scaled by the frequency ratio, $r$, yielding
$\nu_{\rm t}$ phase solutions for the times at which the source was
observed at $\nu_{\rm t}$.

We did not use the phase rates derived by fringe-fitting because each
of those was derived from one half-cycle of 22\,s length, whereas the
phase rates interpolated from two consecutive half-cycles as described
above used two half-cycles separated by 50\,s and therefore have much
better SNR. This requires the phase to change by less than $180^\circ$
in 50\,s, which was the case during our experiment due to sufficiently
short half-cycles. A further advantage of deriving phase rates from
pairs of phase solutions is that each $\nu_{\rm r}$ phase solution is
used both in the determination of the phase rate to the preceding and
the succeeding $\nu_{\rm r}$ solution, causing a smoothing of the
phases with time and reducing the effect of outliers. Finally, the
interpolated phase, phase rate and delay solutions are stored in an
output table together with the $\nu_{\rm t}$ frequency ID. FFSTG
provides an interface to Gnuplot to plot and inspect the input and
output phases. One can select regions of interest and check the
results of the scaling.

The table was imported to AIPS using TBIN and was used to update the
most recent calibration table at the target frequency.\\

\section{Results}
\label{sec:results}

\subsection{43\,GHz}

\begin{figure}[htpb!]
\centering
 \subfigure[\object{NGC\,4261} raw 43\,GHz visibility phases on baselines to
 Los Alamos with only delay calibration applied.]{
 \includegraphics[width=5.5cm, angle=270]{4261_43_tecor_LA.ps}
 \label{fig:4261_43_tecor_LA.ps}}
 \subfigure[\object{NGC\,4261} calibrated 43\,GHz visibility phases on baselines
  to Los Alamos. Calibration used scaled-up phase solutions from
 fringe-fitting with a clean component model at 15\,GHz, but no
  ionospheric correction has been applied.]{
  \includegraphics[width=5.5cm, angle=270]{4261_43_no_tecor_LA.ps}
  \label{fig:no_tecor}}\\
 \subfigure[\object{NGC\,4261} calibrated 43\,GHz visibility phases on baselines to
 Los Alamos. Calibration used ionospheric corrections and scaled-up phase solutions from
 fringe-fitting with a clean component model at 15\,GHz. The residual phase rates
  are much lower than in Fig.~\ref{fig:no_tecor}. ]{
 \includegraphics[width=5.5cm, angle=270]{4261_43_tecor+ffs_LA.ps}
 \label{fig:4261_43_tecor+ffs_LA.ps}}\\
 \caption{Raw and calibrated visibility phases on baselines to Los
 Alamos. Each data point is an average over a half-cycle.}
 \label{fig:results_LA}
\end{figure}

\begin{figure*}[htpb!]
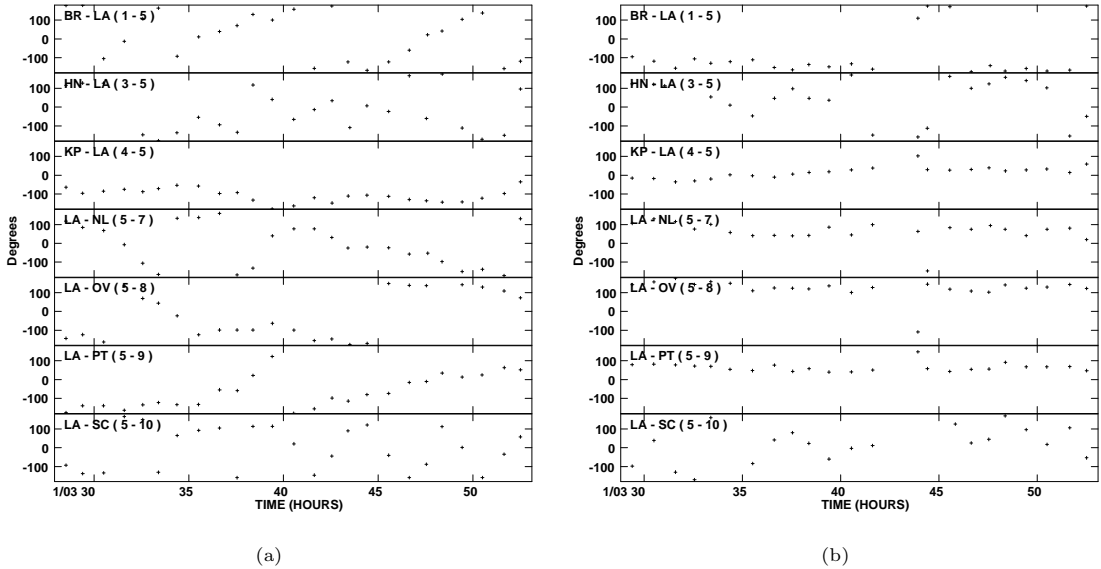

\centering
 \subfigure[]{\includegraphics[width=7cm, angle=270]{4261_43_tecor_LA_short.ps}
 \label{fig:4261_43_tecor_LA_short.ps}}
 \subfigure[]{\includegraphics[width=7cm, angle=270]{4261_43_tecor+ffs_LA_short.ps}
 \label{fig:4261_43_tecor+ffs_LA_short.ps}}\\
 \caption{Enlarged sections of Fig.~\ref{fig:4261_43_tecor_LA.ps} (left panel) 
 and \ref{fig:4261_43_tecor+ffs_LA.ps} (right panel).}
 \label{fig:results_LA_zoom}
\end{figure*}

NGC\,4261 was detected on most baselines at all times after
scaling the 15\,GHz solutions to 43\,GHz. Here, the term ``detected''
means that by inspecting the phase time series by eye one could see
that the phase was not random. The correlated flux densities range
from 30\,mJy on baselines of 800\,M$\lambda$ to 160\,mJy on baselines
of 30\,M$\lambda$.  The 43\,GHz half-cycle average visibilities on
baselines to LA are shown in Figs.~\ref{fig:results_LA} and
\ref{fig:results_LA_zoom}. The short-term fluctuations introduced by
the troposphere are almost perfectly calibrated, but residual phase
drifts remain on longer time-scales, especially at the beginning
(before and during sunset) and at the end of the experiment, when the
source elevation was lower.\\

\subsubsection{Structure functions}

\begin{figure}[htpb!]
 \resizebox{\hsize}{!}{\includegraphics[width=0.95\linewidth,
 clip]{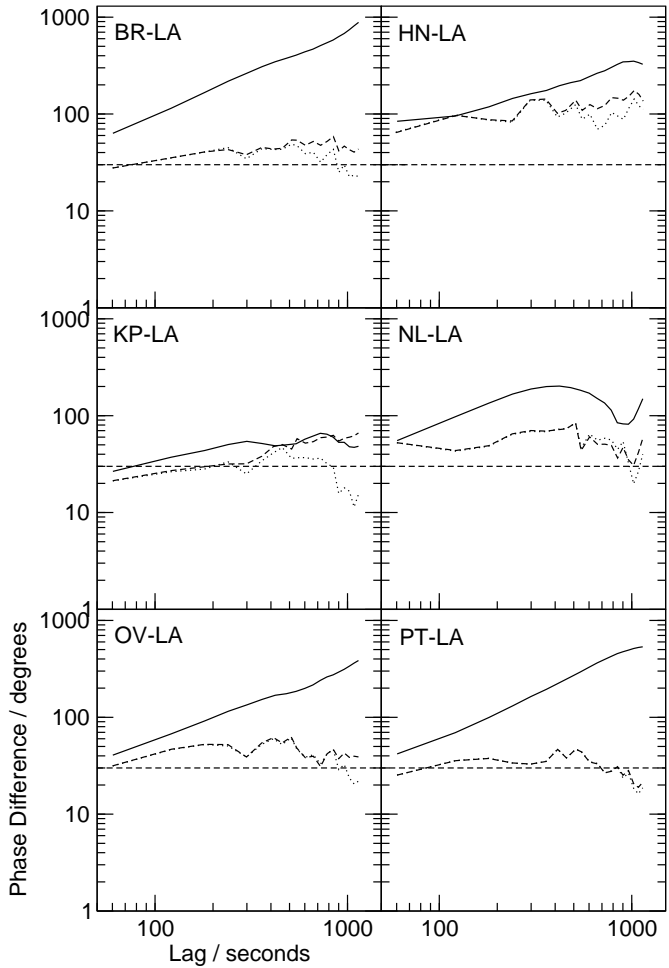}} \caption[Structure functions of
 raw and calibrated 43 GHz data]{Structure functions of the phase time
 series from 25\,min of observation of NGC\,4261 at
 43\,GHz. Solid lines show raw data observed in the middle of the
 experiment with delay calibration only. The data were averaged in
 frequency and over each half-cycle. Dashed lines show the same data
 calibrated with scaled-up phase solutions from fringe-fitting at
 15\,GHz, and dotted lines show the same data after fringe-fitting
 with a 30\,min solution interval. Phase wraps have been removed. The
 horizontal dashed lines indicate the theoretical phase noise. The
 median residual phase noise of calibrated data with residual phase
 rates is $33^{\circ}$ on a 50\,s time-scale, and without residual
 phase rates is $31^{\circ}$. }
\label{fig:struct_func_bm175c}
\end{figure}

Structure functions of the 43\,GHz visibility phases from a 25\,min
scan on NGC\,4261 are shown in
Fig.~\ref{fig:struct_func_bm175c} showing the data at each stage of
calibration. Phase wraps have been removed, allowing phase differences
exceeding $360^\circ$.

The structure functions constructed from the calibrated phase
time-series show a residual phase noise on the shortest time-scales
(50\,s) of $50^{\circ}$ early in the observations and of $33^{\circ}$
in the middle of the experiment (taking the median of the phase noise
of all baselines). The phase noise increases toward longer
time-scales, probably due to errors in the ionospheric
models. Evidence for the long-term phase noise being dominated by
ionosphere is as follows.

First, applying an ionospheric TEC correction significantly improved
the coherence, as one can see by comparing Figs.~\ref{fig:no_tecor}
and \ref{fig:4261_43_tecor+ffs_LA.ps}. Second, the residual phase
errors decreased in the middle of the experiment, when the average
elevation of the source was $58^\circ$. During the first scan at
01:00\,UT, the average source elevation at all antennas was
$32^\circ$, hence the line of sight through the ionosphere was 63\,\%
longer, amplifying errors in the ionospheric model. Also, the
experiment began briefly before sunset at the stations in the
south-west US. This is a time of rapidly changing ionospheric TEC
causing unstable phases early in the experiment on baselines to the
south-western stations.

We fringe-fitted the 43\,GHz data calibrated with fast frequency
switching with a solution interval of 30\,min to remove the residual
long-term phase drifts. The structure functions after applying this
correction are shown as dotted lines in
Fig.~\ref{fig:struct_func_bm175c}; the long-term phase noise is
reduced as expected, and the phase noise on 50\,s time scale is
slightly lower, with a median of $44^{\circ}$ during the beginning and
$31^{\circ}$ in the middle of the experiment.\\

\subsubsection{Expected phase noise}

The expected phase noise in the visibilities calibrated with fast
frequency switching consists of five parts: (1) thermal phase noise at
the reference frequency scaled by the frequency ratio, (2) thermal
phase noise at the target frequency, (3) tropospheric phase changes
during the two integrations, (4) ionospheric scintillations and (5)
errors in the source model at the reference frequency. Estimates of
these contributions are as follows.\\

The $1\,\sigma$ thermal noise on a baseline is given by

\begin{equation}
\Delta S = \frac{1}{\eta_{\rm s}} \times \frac{SEFD}{\sqrt{2\times\Delta\nu\times\tau}}
\end{equation}

(\citealt{Walker1995}), where $\eta_{\rm s}$ is an efficiency factor
(0.69 for the VLBA with 2-bit sampling as we used), $SEFD$ is the
antenna's system equivalent flux density in Jy, $\Delta\nu$ is the
bandwidth in Hz and $\tau$ the integration time in seconds.

The typical measured $SEFD$ for the VLBA antennas was 550\,Jy at
15\,GHz, 1436\,Jy at 43\,GHz and 5170\,Jy at 86\,GHz (at 43\,GHz and
86\,GHz, these numbers are fairly variable between antennas). In our
observations, $\tau$ was 15\,s at 15\,GHz and 21\,s at 43\,GHz and
86\,GHz, and $\Delta\nu$ was 64\,MHz. One therefore expects thermal
noise levels of 18.2\,mJy, 40.1\,mJy and 145\,mJy at 15\,GHz, 43\,GHz
and 86\,GHz, respectively. At 15\,GHz, NGC\,4261 has a
correlated flux density, $S$, between $60\,{\rm mJy}$ and $300\,{\rm
mJy}$, depending on baseline length, so we adopt either $80\,{\rm
mJy}$, representative of long baselines, or $200\,{\rm mJy}$,
representative of short baselines (the results of which we give in
brackets).\\

\paragraph{(1)} On a single baseline, the expected SNR of a detection
at 15\,GHz is $S/\Delta S=80\,{\rm mJy}/18.2\,{\rm mJy}=4.61$ (11.0)
when averaging over the band. Fringe-fitting, however, uses all
baselines to one particular antenna to derive a phase correction, and
so the SNR is increased by $\sqrt{N}$, where $N$ is the number of
baselines. $N\approx8$, so the SNR of a detection at 15\,GHz increases
to 12.4 (31.1). We have converted the SNR into a visibility phase rms
using the following approach. Let us assume that the phase of the true
visibility is 0. The probability distribution of the measured
visibility phase as a function of SNR, $p(\phi)$, is given in
Eq.~6.63b in \cite{Thompson2001}. The square of the phase rms derived
from that probability is:

\begin{equation}
\sigma^2 = \frac{\int\phi^2 p(\phi)d\phi}{\int p(\phi)d\phi}
\label{eq:rms}
\end{equation}

We have evaluated the integral numerically and find a phase rms of
$4.61^{\circ}$ ($1.84^{\circ}$). This phase noise is scaled by the
frequency ratio to $13.8^{\circ}$ ($5.53^{\circ}$) at 43\,GHz and
$27.6^{\circ}$ ($11.1^{\circ}$) at 86\,GHz.

\paragraph{(2)} Assuming that NGC\,4261 has a compact flux density of
100\,mJy at both 43\,GHz and 86\,GHz, the thermal noise contributions
according to Eq.~\ref{eq:rms} are $26.3^{\circ}$ and $76.0^{\circ}$,
respectively. Adding those in quadrature to the scaled-up thermal rms
phase noise from 15\,GHz yields $29.7^{\circ}$ ($26.9^{\circ}$) at
43\,GHz and $80.9^{\circ}$ ($76.8^{\circ}$) at 86\,GHz.

\paragraph{(3)} We estimated the tropospheric phase noise within the
switching cycle time using structure functions of \object{3C\,273}. On
\object{3C\,273}, the thermal noise contributions at 15\,GHz and
43\,GHz are $0.04^{\circ}$ and $0.4^{\circ}$, respectively, so that
the visibility phases are essentially free of thermal noise and any
phase changes during and between the half-cycles are due to changes in
the troposphere. We found the median rms phase noise on a 50\,s
time-scale after fringe-fitting with a 30\,min solution interval to
remove the residual long-term phase drift to be $13.3^{\circ}$ at
43\,GHz, or $26.6^{\circ}$ at 86\,GHz.\\

The amplitudes of (4) (ionospheric scinitillations) were found to be
less than three radians at 378\,MHz on time-scales up to 40\,s by
\cite{Yeh1982}. When scaled to 15\,GHz, they contribute only
$0.1^\circ$ on time-scales of 40\,s, and even less at the higher
frequencies, so their contribution remains miniscule and is no longer
considered. The contribution of (5) (errors in the source model) is
also not significant. First, these errors do not contribute a
noise-like component but rather long-term phase errors similar to
errors in the TEC model. Second, phase corrections at 15\,GHz to
account for the source structure were at most $50^\circ$ in 6\,h, or
$3.5^\circ$ in 25\,min, and only on the longest baselines. This means
that introducing a source model at all required changes of order a few
degrees per 30\,min. Errors in the model, however, are expected to be
much smaller because the sampling of the $(u,v)$ plane at 15\,GHz was
excellent. They should barely exceed $1^\circ$ at 15\,GHz, or
$3^\circ$ at 43\,GHz. This is much less than all other source of phase
noise, and can therefore be neglected.

Adding the first three non-negligible noise components in quadrature
yields $32.6^{\circ}$ ($30.0^{\circ}$) at 43\,GHz and is in excellent
agreement with the measured rms phase noise of $31^{\circ}$.\\

\subsubsection{Coherence}

\begin{figure*}[ht!]
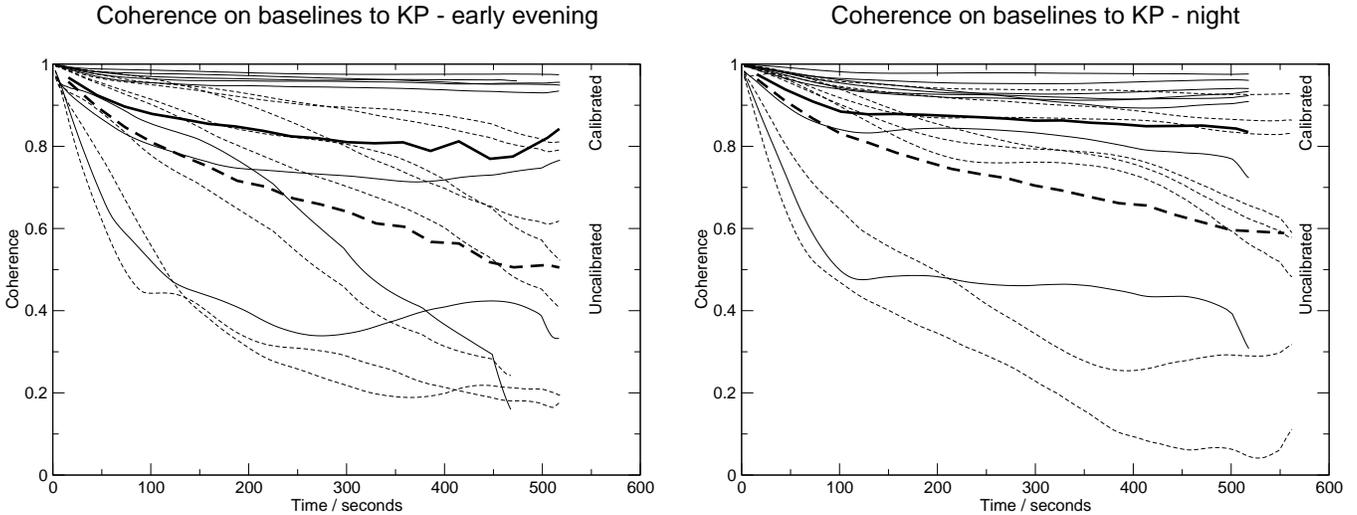

\centering
 \hbox{ \includegraphics[width=0.48\linewidth,
 clip]{KP_3h14.eps} \hspace*{3mm}
 \includegraphics[width=0.48\linewidth, clip]{KP_6h47.eps} }
 \caption{Coherence diagrams showing the improvement achieved at
 43\,GHz using fast frequency switching.  The vector sum over
 progressively longer time intervals has been computed on all
 baselines and has been normalized by the scalar sum of the vectors,
 after setting the phases to zero. Bold lines show the average over
 all baselines in bins of 28\,s, i.e., the length of one half-cycle.
 The coherence was analysed using visibilities on baselines to Kitt
 Peak, which had moderate to good (but not excellent) weather
 conditions, and hence is representative for the array performance
 during the observing run. Shown are \object{3C\,273} data observed in
 the beginning (centred on 03:20\,UT, left panel) and the middle
 (centred on 06:50\,UT, right panel) of the experiment at 43\,GHz with
 interleaved 15\,GHz observations for calibration. 43\,GHz data from
 10 min scans with (solid lines) and without (dashed lines) phase
 calibration from 15\,GHz have been averaged in frequency. As 3C\,273
 is strong, the data are essentially free of thermal noise and show
 the coherence improvement with fast frequency switching.}
\label{fig:coherence}
\end{figure*}

Another way of illustrating the performance of the corrections is with
coherence diagrams (Fig.~\ref{fig:coherence}). For the plots shown,
\object{3C\,273} 43\,GHz data from all baselines to KP have been
used. All data were calibrated using fast frequency switching (no
subsequent fringe-fitting), and were averaged over the band and in
time to obtain one visibility per baseline and half-cycle. The vector
average of these visibilities has been computed over progressively
longer time intervals and normalized by the scalar average.

The left panels show data from a 25\,min scan centred on 03:20\,UT and
the right panels show data from a 25\,min scan centred on 06:50\,UT.
Bold lines show the average over all baselines. The coherence achieved
in the middle of the experiment with fast frequency switching is a
little better than that achieved during the beginning of the
observations.

\subsubsection{Stability of the phase offset}

After correction of the ionopheric phase errors, the visibility phases
on the LA-PT baseline were stable to $<$1\,rad over 10\,h
(Fig.~\ref{fig:4261_43_tecor+ffs_LA.ps}). This indicates that the
instrumental phase offset, $\Delta\Phi$, was also stable within the
same limits, and it appears to be sufficient to determine $\Delta\Phi$
a few times throughout the experiment.

\subsubsection{Imaging}
\label{sec:imaging}

\begin{figure}[ht!]
\centering
 \includegraphics[width=\linewidth]{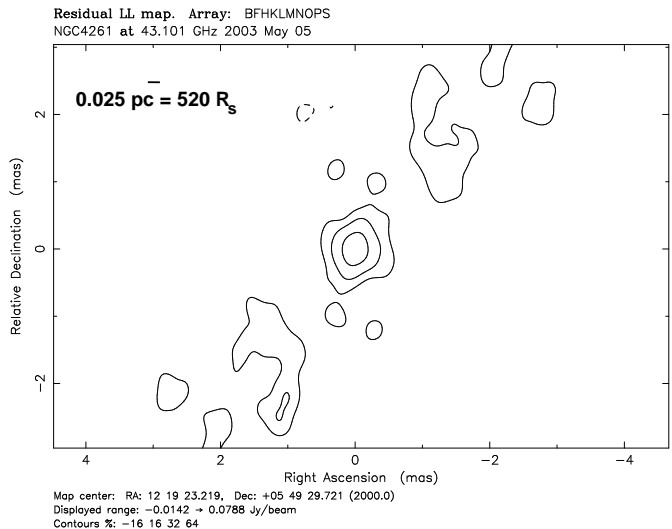}
 \caption[NGC\,4261 43\,GHz dirty image]{Naturally weighted,
 full-resolution dirty image of NGC\,4261 at 43\,GHz, calibrated
 with scaled-up phase solutions from 15\,GHz. Fringe-fitting has been
 used to solve for one residual phase and rate solution per 25\,min
 scan before exporting the data to Difmap. No further self-calibration
 has been applied. The image noise is 4.4\,mJy\,beam$^{-1}$ and the
 dynamic range is 18:1. The synthesized beam size is
 $0.38\,{\rm mas}\times0.18\,{\rm mas}$. The bar in the upper left corner shows
 the size of the minor axis.} \label{fig:NGC4261_43GHz_FFS_raw}
\end{figure}

\begin{figure}[ht!]
\centering
 \includegraphics[width=\linewidth]{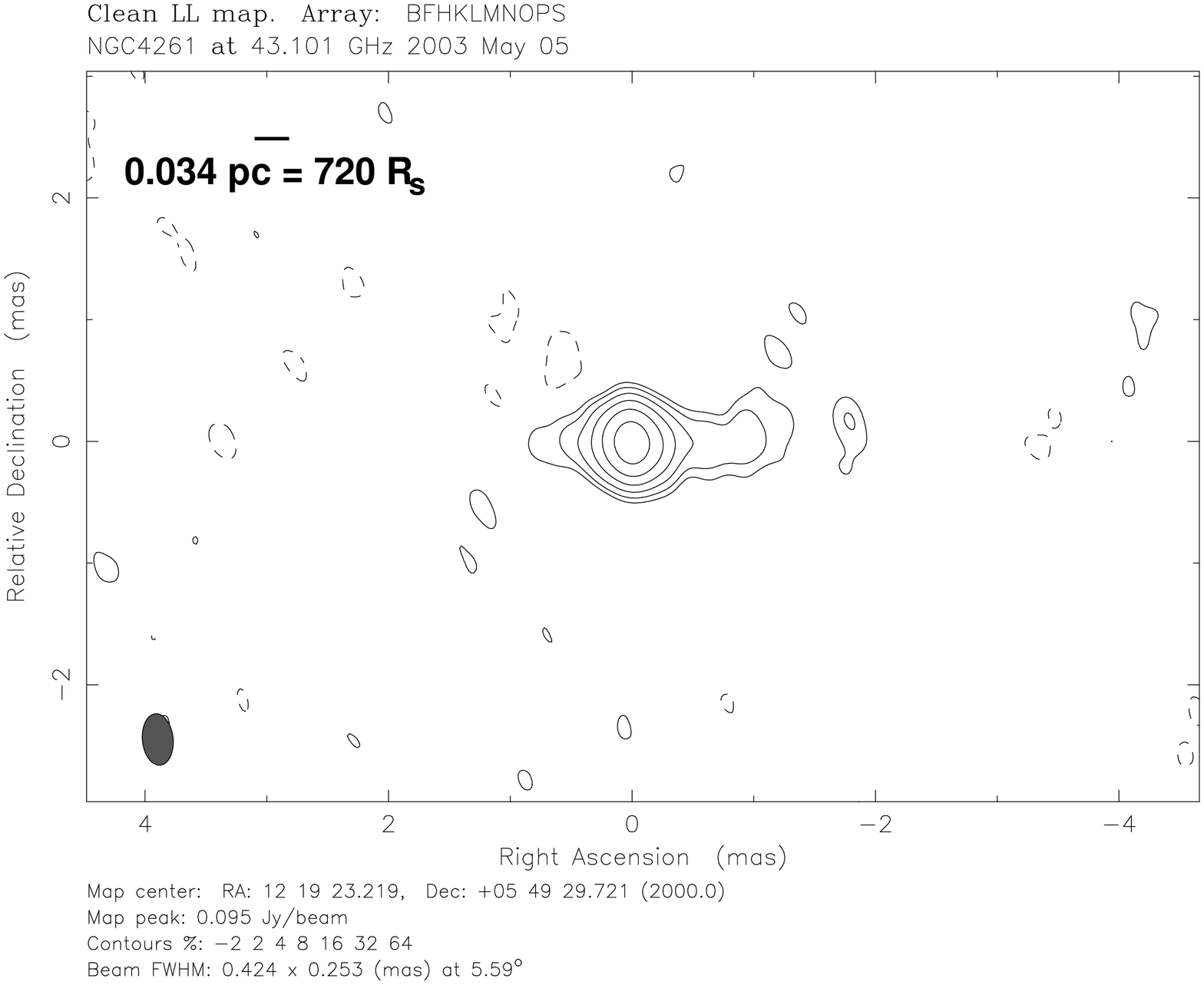}
 \caption[NGC\,4261 43\,GHz clean image]{Data and imaging
 parameters as in Fig.~\ref{fig:NGC4261_43GHz_FFS_raw}, but several
 cycles of phase self-calibration with a solution interval of 30\,s
 and one cycle of amplitude self-calibration with a solution interval
 of 12\,h have been applied. The final image was cleaned and has an
 rms noise of 0.78\,mJy\,beam$^{-1}$ and a dynamic range of
 122:1. The synthesized beam size is $0.42\,{\rm mas}\times0.25\,{\rm
 mas}$. The bar in the upper left corner shows the size of the minor
 axis.}  \label{fig:NGC4261_43GHz_FFS}
\end{figure}

Before making an image from the 43\,GHz visibilities calibrated with
fast frequency switching, the residual phase offsets and phase rates
were removed using fringe-fitting with a solution interval of 30\,min
so that one solution per scan was obtained. The resulting dirty image
(Fig.~\ref{fig:NGC4261_43GHz_FFS_raw}) has a peak flux density of
$79\,{\rm mJy\,beam^{-1}}$ and an rms noise of $4.4\,{\rm
mJy\,beam^{-1}}$, yielding a dynamic range of 18:1. After several
cycles of phase self-calibration with a solution interval of 30\,s and
one cycle of amplitude self-calibration with a solution interval of
12\,h, the peak flux density was $95\,{\rm mJy\,beam^{-1}}$ and the
rms noise was $0.78\,{\rm mJy\,beam^{-1}}$, so the dynamic range
improved to 122:1 (Fig.~\ref{fig:NGC4261_43GHz_FFS}). The theoretical
rms noise at 43\,GHz was expected to be $0.46\,{\rm
mJy\,beam^{-1}}$. Unfortunately, the ionospheric phase errors and the
need to fringe-fit to calibrate them did not allow a core shift
measurement to be made.

Our final image has a synthesized beam size of $0.42\,{\rm
mas}\times0.25\,{\rm mas}$, corresponding to a linear resolution of
$720\,R_{\rm s}$. This is about 50\,\% larger than achieved by
\cite{Jones2000} with conventional, non-phase-referencing 43\,GHz VLBA
observations.  The difference in resolution is due to two
factors. First, we made only few detections on baselines to Mauna
Kea. This is because Mauna Kea was observing at predominantly low
elevations, hence suffering more from ionospheric phase
errors. Second, we used natural weighting to increase sensitivity to
extended structures, whereas \cite{Jones2000} used uniform weighting
for increased resolution.

Notwithstanding these differences, our 43\,GHz image is very similar
to that published as Fig.~5 in \cite{Jones2000}. Both observations
show a dominant core and a jet extending 1.4\,mas to the west, with
indications of emission at 2\,mas west of the core. Both images
indicate the presence of a counterjet extending up to 1\,mas east of
the core. However, \cite{Jones2000} find a peak flux density of
141\,mJy\,beam$^{-1}$, whereas we find a peak flux density of
95\,mJy\,beam$^{-1}$. The observations were separated by 5.7\,yr, and
source variability may have caused some (or all) of the
difference. Furthermore, changing tropospheric opacity during either
observation may have affected the peak flux density. In our image, the
eastern jet looks smooth, whereas it shows components in the
observation of \cite{Jones2000}. This probably arises from the
different weighting schemes and image fidelity limits, rather than
from a real change of jet character.

\subsection{86\,GHz}
\label{sec:86GHz_results}

\begin{figure}[htpb]
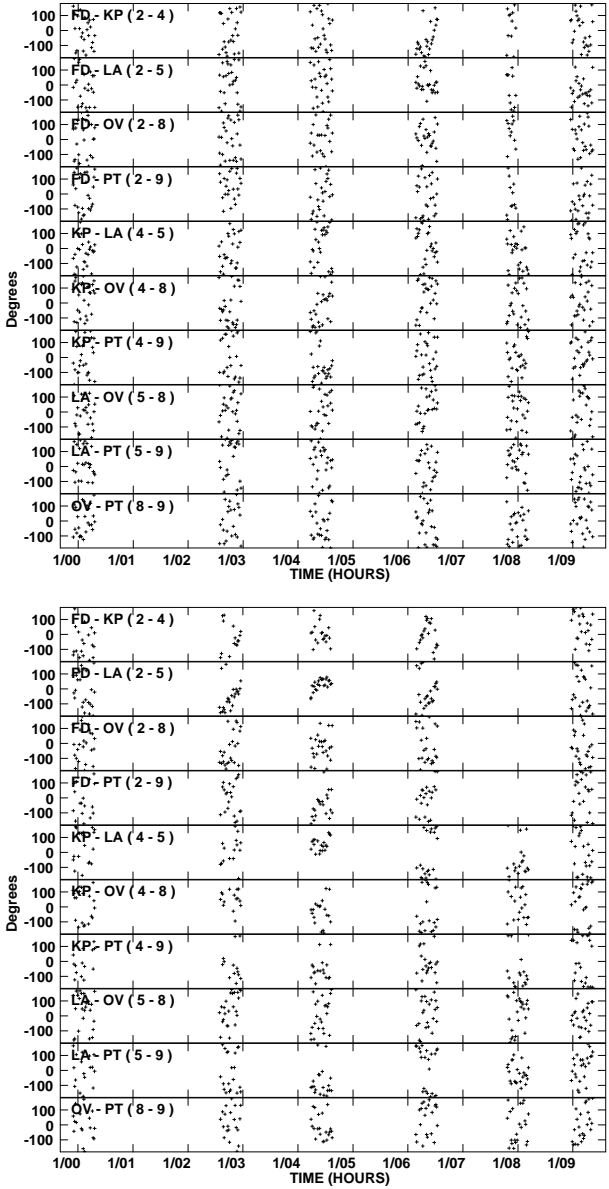

\begin{minipage}[b]{8cm}
 \includegraphics[width=8cm, angle=270]{4261_86_CL11_tecor.ps}
 \includegraphics[width=8cm, angle=270]{4261_86_CL17_tecor+ffs.ps}
\end{minipage}
\caption[86\,GHz visibility phases]{86\,GHz visibility phases. {\it
Top:} NGC\,4261 raw 86\,GHz visibility phases on baselines
among Fort Davis, Kitt Peak, Los Alamos, Owens Valley and Pie Town
with only delay calibration applied. {\it Bottom:} 86\,GHz visibility
phases on baselines among Fort Davis, Kitt Peak, Los Alamos, Owens
Valley and Pie Town. Calibration has been done with scaled-up phase
solutions from fringe-fitting at 15\,GHz using a clean component
model. Good detections were made during almost every 25\,min scan
observed at night between 2:00\,UT and 7:00\,UT, when the source
elevation was high and ionospheric effects lower.}
\label{fig:4261_86}
\end{figure}

\begin{figure}
 \centering
 \includegraphics[width=0.9\linewidth, clip]{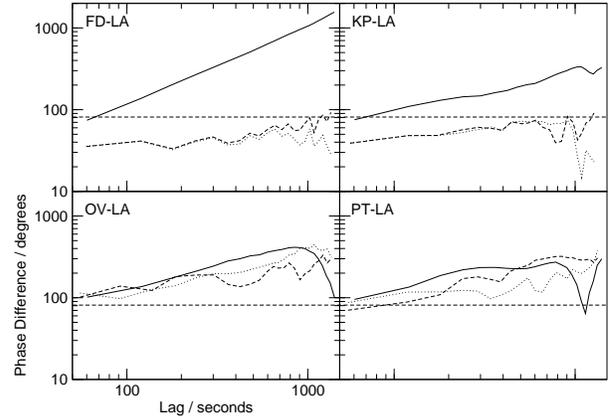}
 \caption[Phase structure functions of a 25\,min scan at
 86\,GHz]{Structure functions of a 25\,min scan at 86\,GHz, showing
 the same stages of calibration as in
 Fig.~\ref{fig:struct_func_bm175c}. Note that the expected
 noise level, indicated by the horizontal dashed lines, is a guideline
 only. It depends on the correlated source flux density which varies
 with baseline length and orientation (see Section~\ref{sec:86GHz_results}).}
\label{fig:struct_funct_86GHz_bm175c}
\end{figure}

\begin{figure}[ht!]
 \includegraphics[width=\linewidth, clip]{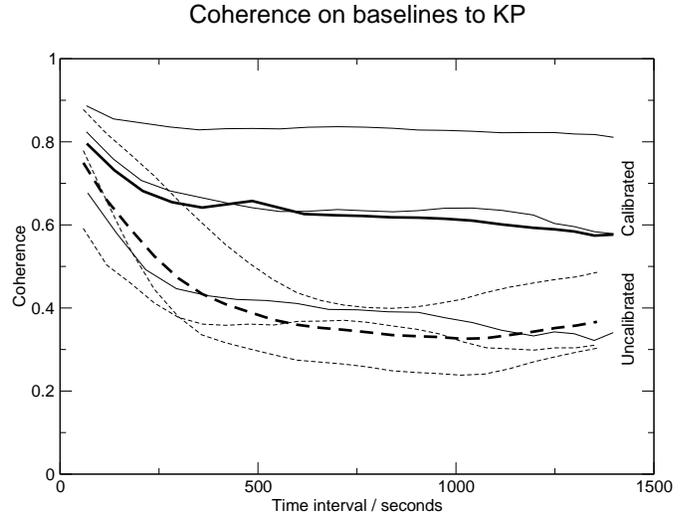}
 \caption{The coherence improvement at 86\,GHz with fast frequency
 switching on baselines to KP, obtained in the same way as for
 Fig.~\ref{fig:coherence}. However, we did not observe any calibrator long
 enough to generate coherence plots, so the diagram was constructed
 from NGC\,4261 data. The coherence therefore is
 significantly lowered by thermal noise.}
\label{fig:coherence_86}
\end{figure}

\begin{figure}[htpb!]
 \includegraphics[width=9cm]{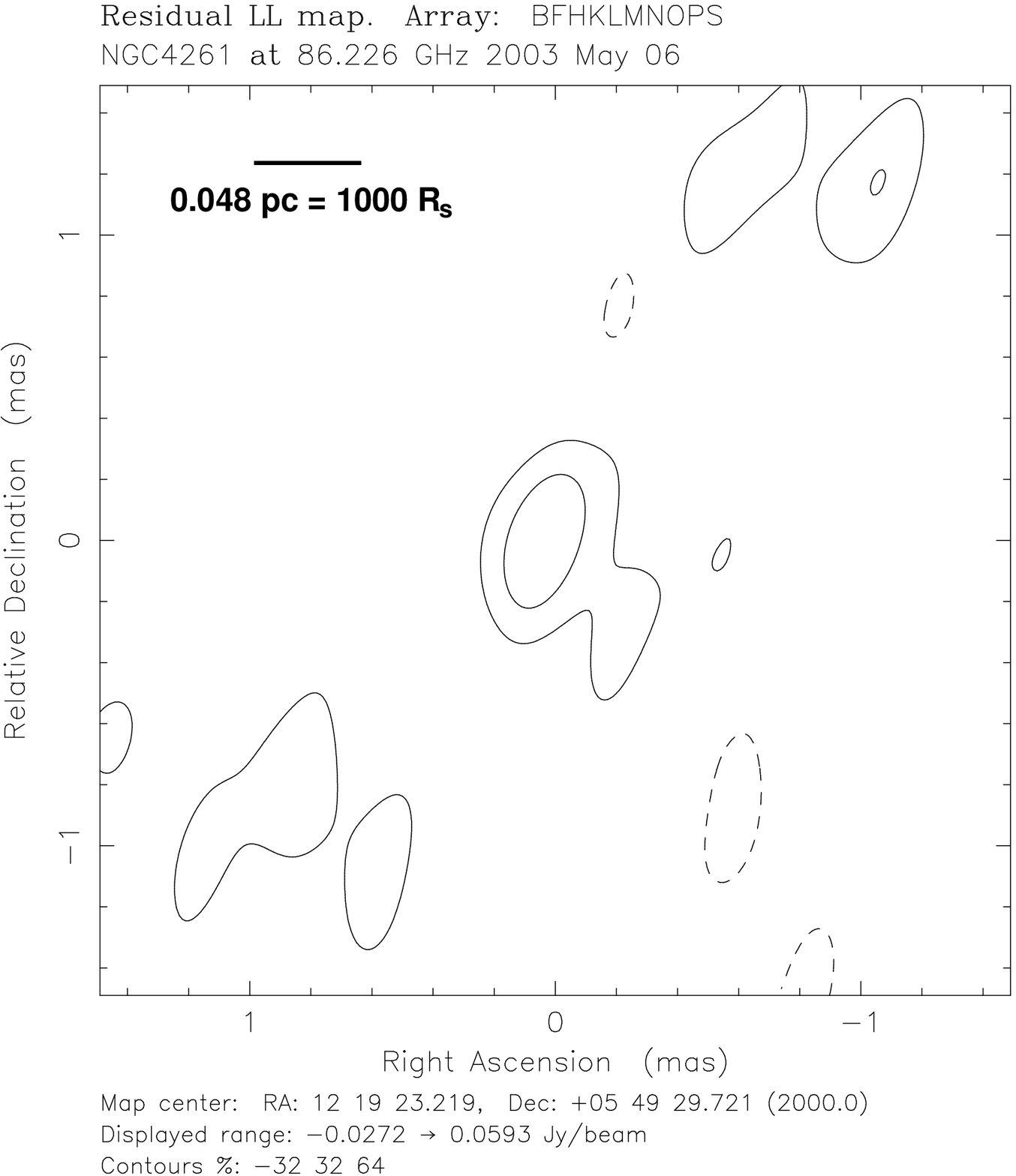}
\caption{Naturally weighted, full-resolution dirty
image of NGC\,4261 at 86\,GHz, calibrated with scaled-up
phase solutions from 15\,GHz. Fringe-fitting has been used to solve
for one residual phase and rate solution per 25\,min scan before
exporting the data to Difmap. No further self-calibration and no
deconvolution has been applied. The image noise is
8.4\,mJy\,beam$^{-1}$ and the dynamic range is 7:1. The synthesized
beam has a diameter of $0.35\,{\rm mas}\times0.54\,{\rm mas}$ in P.A. $11^\circ$
and corresponds to a linear resolution of 0.048\,pc or $1000\,R_{\rm
s}$. The bar in the upper left corner shows the size of the minor
axis.}
\label{fig:NGC4261_86GHz}
\end{figure}

\begin{figure}[htpb!]
 \includegraphics[width=9cm]{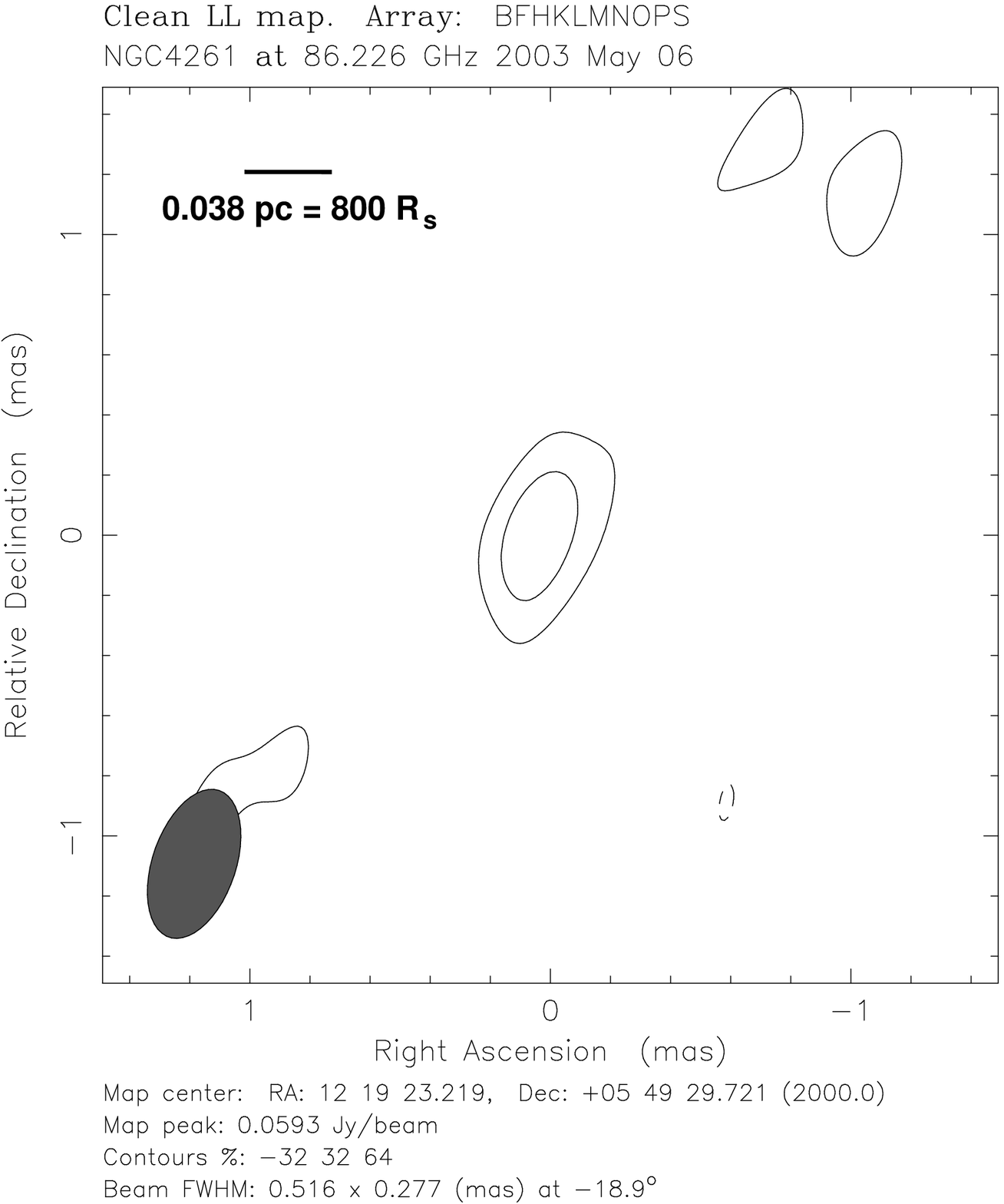}
\caption{Naturally weighted, full-resolution clean
image of NGC\,4261 at 86\,GHz using the same data and
calibration as in Fig.~\ref{fig:NGC4261_86GHz}. No further
self-calibration has been applied. The image noise is
7.5\,mJy\,beam$^{-1}$ and the dynamic range is 8:1. The synthesized
beam has a diameter of $0.28\,{\rm mas}\times0.52\,{\rm mas}$ in
P.A. $19^\circ$ and corresponds to a linear resolution of 0.038\,pc or
$800\,R_{\rm s}$.}
\label{fig:NGC4261_86_CL17_clean}
\end{figure}

Following the same data reduction path as for the 43\,GHz data, we
obtained good detections of NGC\,4261 at 86\,GHz on baselines
among the four stations FD, KP, LA, and PT and only weak detections on
baselines to NL, OV and MK. Again, a ``detection'' means that in a
plot of visibility phase with time as in Fig.~\ref{fig:4261_86}, one
can see that the phases cluster and are not random. 

Uncorrected and corrected visibility phases are plotted in
Fig.~\ref{fig:4261_86}. The improvement in coherence is clearly
visible as a pronounced clustering of points in the lower panel,
compared to quasi-random phases in the upper panel. Structure
functions and coherence plots from a 25\,min scan of
NGC\,4261 are shown in
Figs.~\ref{fig:struct_funct_86GHz_bm175c} and
\ref{fig:coherence_86}. The coherence improvement is especially
pronounced on the FD-LA baseline (upper left panel in
Fig.~\ref{fig:struct_funct_86GHz_bm175c}), whereas the improvement on
the PT-LA and OV-LA baselines is small. The coherence plot in
Fig.~\ref{fig:coherence_86} shows that the coherence on baselines to
KP increased from 35\,\% to almost 60\,\% on time-scales of 20\,min.

A dirty image is shown in Fig.~\ref{fig:NGC4261_86GHz} and a cleaned
image in Fig.~\ref{fig:NGC4261_86_CL17_clean}.  The peak flux density
is 59.3\,mJy. This is the first detection of
NGC\,4261 with VLBI at 86\,GHz. It is also probably the
weakest continuum object ever detected with VLBI at this
frequency. The next brightest detection of which we are aware is
85\,mJy in the 86\,GHz image of $1308+328$ by
\cite{Porcas2002}, using conventional phase referencing to a
calibrator $14.3^{\prime}$ away. With only delay calibration applied,
the median rms phase noise of the best 25\,min scan is $104^{\circ}$,
after applying the scaled 15\,GHz phase solutions is $70^{\circ}$ and
after fringe-fitting with a 30\,min solution interval is
$80^{\circ}$. The increase in rms phase noise after removal of phase
rates is unexpected and has an unknown cause.

The expected thermal phase noise is the quadrature sum of the
scaled-up 15\,GHz noise and the 86\,GHz noise, the former of which we
estimated to be $27.6^{\circ}$ on long and $11.1^{\circ}$ on short
baselines, and the latter, assuming a correlated flux density of
100\,mJy and using Eq.~\ref{eq:rms}, is $76.0^{\circ}$. The quadrature
sum is $80.9^{\circ}$ on long and $76.8^{\circ}$ on short baselines,
and adding the tropospheric phase errors of $26.6^{\circ}$ is
$85.1^{\circ}$ and $81.3^{\circ}$, in agreement with the measured
noise levels.

Note that these numbers provide a guideline only. The largest
contribution to the expected 86\,GHz noise is the $76.0^{\circ}$ from
thermal noise in the 86\,GHz receivers. This number depends on the
source flux density which is a strong function of baseline length and
orientation. For example, the average visibility amplitude on the
relatively short FD-LA baseline was 180\,mJy at 4:30\,UT (not 100\,mJy
as adopted above), reducing the 86\,GHz thermal noise to
$55.2^\circ$. The expected total phase noise, indicated by the dashed
horizontal line in the upper left panel in
Fig.~\ref{fig:struct_funct_86GHz_bm175c}, then drops to $62.3^\circ$.

The rms noise in the final 86\,GHz image is $7.4\,{\rm
mJy\,beam^{-1}}$, compared to an expected thermal noise of $3.7\,{\rm
mJy\,beam^{-1}}$. The dynamic range in the image is 7:1.

\subsection{Discussion of the 86\,GHz image}

The final 86\,GHz image of NGC\,4261 in
Fig.~\ref{fig:NGC4261_86_CL17_clean} has a synthesized beam size of
$0.28\,{\rm mas}\times0.52\,{\rm mas}$ in P.A. $19^\circ$, which is
about 12\,\% larger than the clean beam of the 43\,GHz image. This is
mostly due to the lack of detections on long baselines and the phase
noise arising from low elevations at Mauna Kea (see
Section~\ref{sec:imaging}) was larger by a factor of two at 86\,GHz
than at 43\,GHz due to the higher frequency. The VLBA station at
Hancock was being commissioned at 86\,GHz and the station at Brewester
was not equipped at 86\,GHz; both would have contributed long
baselines at this frequency. Also, the complete loss of 43\,GHz data
at Fort Davis resulted in poorer sampling of short $(u,v)$ spacings at
43\,GHz but not at 86\,GHz, so there is an emphasis on long baselines
at 43\,GHz and on short baselines at 86\,GHz. The resolution at
86\,GHz corresponds to 0.038\,pc, or $800\,R_{\rm s}$, at the distance
of NGC\,4261.

The image displays a point source only; further self-calibration of
the data did not reliably converge on any extended structure. The
point source is probably the brightest component seen in the 43\,GHz
image, which probably corresponds to the jet base. Our observations
therefore did not allow us to resolve the jet collimation
region. However, we are confident that the suggested improvements to
the observing strategy to calibrate ionospheric phase errors can
significantly improve the detection rates on long baselines and hence
increase the resolution.

\section{Benefits of fast frequency switching}
\label{sec:benefits}

The primary use of fast frequency switching is to detect sources that
are too weak for self-calibration at the target frequency within the
atmospheric coherence time, but can be reliably detected at a lower
frequency. The $5\,\sigma$ detection limit of the VLBA at 43\,GHz
within 120\,s, with 64\,MHz bandwidth and 2-bit sampling, and
neglecting coherence loss due to tropospheric phase changes, is
84.0\,mJy (i.e., a thermal noise level of 16.8\,mJy on a single
baseline). The $1\,\sigma$ noise level using fast frequency switching
with half-cycle times of 22\,s at 15\,GHz and 28\,s at 43\,GHz
(yielding net integration times of 15\,s and 21\,s, respectively),
after one cycle is 67.7\,mJy, of which 40.1\,mJy is thermal noise in
the raw 43\,GHz visibility, 54.6\,mJy is thermal noise in the phase
solutions after scaling from 15\,GHz, and neglecting noise due to
tropospheric phase changes. The $1\,\sigma$ noise level after 120\,s
of fast frequency switching (i.e., 2.4 cycles) is 43.7\,mJy. The noise
level of fast frequency switching reaches that of a conventional
120\,s integration at 43\,GHz of 16.8\,mJy after 812\,s (13.5\,min).
Any longer integration with fast frequency switching then yields
detection thresholds that are out of reach with conventional methods.

Furthermore, the proposed application of fast frequency switching to
detect core shifts in AGN can supplement jet physics with
observational constraints which are otherwise difficult to obtain.\\

Fast frequency switching is limited by the source strength at the
reference frequency, and 100\,mJy at 15\,GHz is, from our experience,
a reasonable minimum required for a successful observation. Many
sources exist that meet this criterion. The 130 sources in the 2\,cm
VLBI observations by \cite{Kellermann1998}, for example, have peak
flux densities of 100\,mJy\,beam$^{-1}$ or more; all of them can
be observed at 86\,GHz using fast frequency switching.

Another estimate of the number of new sources made accessible to
86\,GHz VLBI by fast frequency switching can be made by means of the
$\log N-\log S$ relation. Let us assume that the sources observable
with conventional 86\,GHz VLBI are distributed homogeneously in an
Euclidean universe, so that lowering the detection limit does not
yield the detection of a new population. The number of sources above a
given flux-density limit, $S$, then increases as $S^{-1.5}$. We
further make the conservative assumption that these sources have
$\alpha=0$ between 15\,GHz and 86\,GHz, because they are compact.

The current $5\,\sigma$ detection limit of the VLBA at 86\,GHz is
605\,mJy (five times the thermal noise of a 30\,s integration with
64\,MHz bandwidth and 2-bit sampling on a single baseline). In
fringe-fitting, all $N$ antennas are used to derive a phase solution,
so the noise level is reduced by $\sqrt{N}$, and with $N=8$, the
$5\,\sigma$ detection limit is 214\,mJy. Fast frequency switching
allows one to observe any compact source with $S_{\rm 86\,GHz}\ge
100\,{\rm mJy}$, a factor of 2.14 fainter than with conventional
methods, and one expects 3.1 times as many sources brighter than
100\,mJy than there are brighter than 214\,mJy, following the $\log
N-\log S$ relation. A little less conservative estimate would be to
allow for spectral indices as steep as $\alpha=-0.3$, so that sources
with $S_{\rm 15\,GHz}=100\,{\rm mJy}$ have $S_{\rm 86\,GHz}>60\,{\rm
mJy}$. The number of observable sources would then increase by a
factor of 6.7 times the number accessible to conventional techniques.

The absolute number of presently observable sources is much more
difficult to determine due to selection effects in the various radio
source catalogues. However, we can make a simple estimate using the
recent 86\,GHz VLBI survey by \cite{Lobanov2002}. They aimed at
observing more than 100 sources selected from the literature that had
an expected compact flux density of $S_{\rm 86\,GHz}>300\,{\rm
mJy}$. Applying the fast frequency switching sensitivity estimates
then yields between 520 and 1100 sources observable with the VLBA at
86\,GHz.


\section{Summary}
\label{sec:summary}

\begin{enumerate}
  
\item Fast frequency switching can be used to calibrate tropospheric
  phase fluctuations if the switching cycle time is shorter than the
  atmospheric coherence time. The accuracy is limited by tropospheric
  phase changes between and during the reference frequency
  integrations.
  
\item Insufficient knowledge about the ionosphere's total electron
  content (TEC) has prevented calibration of the inter-band phase
  offset and hence it was not possible to make a pure phase-referenced
  image without using self-calibration. This also prevented the
  detection of a core shift. Current global TEC models derived from
  GPS data have errors that are too large to sufficiently calibrate
  the ionospheric component of the phase changes. A possible solution
  is to insert frequent (every 10\,min) scans in the 1.4\,GHz band
  with widely separated IF frequencies to derive the ionospheric delay
  on the line of sight to the target source.

\item On the most stable baseline between LA and PT, the instrumental
  phase offset was stable to $<$1\,rad over 10\,h. It therefore seems
  to be sufficient to determine the offset a few times throughout the
  experiment.
  
\item Truly simultaneous observations at two bands would reduce the
  residual phase errors even further because tropospheric phase
  changes within the switching cycle time would be entirely removed.

\item At 86\,GHz, NGC\,4261 was detected as a point source with
 59.3\,mJy\,beam$^{-1}$ flux density and so is the weakest source ever
 detected with VLBI at that frequency. Our observation at 43\,GHz
 yielded a resolution corresponding to $720\,R_{\rm s}$ and hence
 ranks among the highest resolutions achieved in any AGN in terms of
 Schwarzschild radii.

\end{enumerate}

\begin{acknowledgements}

We thank R.~W. Porcas for useful discussions throughout the project
and for his careful review of the manuscript. His critique resulted in
substantial improvements.

\end{acknowledgements}

\bibliography{refs}

\end{document}